\documentstyle[12pt]{article}
\def\beq{\begin{equation}}
\def\eeq{\end{equation}}
\def\bea{\begin{eqnarray}}
\def\eea{\end{eqnarray}}
\def\bq{\begin{quote}}
\def\eq{\end{quote}}

\def\nnb{\nonumber}
\def\ga{\left(}
\def\dr{\right)}
\def\aga{\left\{}
\def\adr{\right\}}

\def\nnb{\nonumber}
\def\la{\langle}
\def\ra{\rangle}
\def\nin{\noindent}
\begin{document}
\topmargin -1.5cm
\oddsidemargin +0.2cm
\evensidemargin -1.0cm
\pagestyle{empty}
\begin{flushright}
{CERN-TH.7441/94}\\
\end{flushright}
\vspace*{5mm}
\begin{center}
\section*{
\boldmath{$K^0$}-\boldmath{$\overline{K}^0$} mixing
and the CKM parameters \boldmath{$(\rho,\eta)$} \\
from the Laplace sum rules}
\vspace*{0.5cm}
{\bf S. Narison} \\
\vspace{0.3cm}
Theoretical Physics Division, CERN\\
CH - 1211 Geneva 23, Switzerland\\
and\\
Laboratoire de Physique Math\'ematique\\
Universit\'e de Montpellier II\\
Place Eug\`ene Bataillon\\
34095 - Montpellier Cedex 05, France\\
\vspace*{1.5cm}
{\bf Abstract} \\ \end{center}
\vspace*{2mm}
\noindent
Using the
Laplace sum rule (LSR)
approach, which is less affected by the contribution
of the higher mass hadronic states than the Finite Energy Sum Rule
(FESR),
we test the reliability of the existing estimate of the $K^0$-$\overline
{K}^0$ mixing parameter from the four-quark two-point correlator.
We obtain, for the renormalization group invariant
$B$-parameter $\Big[ f_K/(1.2f_\pi)\Big]^2 \hat {B}_K$, the upper bound:
0.83 and the $best$ estimate: $0.55 \pm 0.09$.
Combining the previous estimate with the updated value
of $f_B\sqrt{B_B}=(1.49\pm 0.14)f_\pi$ obtained from the same LSR method,
one can deduce the $best$ fitted values
$(\rho,\eta)\approx (0.09,0.41)$
of the CKM parameters.
\vspace*{2.0cm}

\begin{flushleft}
CERN-TH.7441/94 \\
September 1994
\end{flushleft}
\vfill\eject
\setcounter{page}{1}
 \pagestyle{plain}
\section{Introduction} \par
 A reliable estimate of the $B_K$ parameter is of prime importance
for understanding CP-violation. In a series of papers Pich and de Rafael
and coworkers \cite{PICH,DOM}
have estimated its value using the Finite Energy
Sum Rule (FESR), from
the two-point correlator:
\beq
\Psi(q^2) = i \int d^4x e^{iqx} \la 0|{\cal T}
{\cal O}_{\Delta S=2}(x)
{\cal O}^{\dagger}_{\Delta S=2}(0) |0 \ra,
\eeq
 associated to the $\Delta S=2$ four-quark
operator:
\beq
{\cal O}_{\Delta S=2}\equiv \bar s \gamma^\mu L d\bar s \gamma_\mu L s,
\eeq
where $L \equiv (1-\gamma_5)/2$, while the $B_K$-parameter is defined as:
\beq
\la \overline{K}^0| {\cal O}_{\Delta S=2}|K^0 \ra =
\frac{4}{3}f^2_KM^2_KB_K,
\eeq
where $f_K\simeq 1.2 f_\pi$ is the kaon decay constant
($f_\pi=93.3$ MeV). The renormalization group invariant (RGI)
$B$-parameter is defined as
\cite{PICH}:
\beq
\hat {B}_K = B_K \ga \alpha_s (\mu^2) \dr^{-2/9},
\eeq
and its latest improved value from FESR is \cite{DOM}:
\beq
\hat {B}_K= 0.39 \pm 0.10
\eeq
 for $f_K \simeq 1.28f_\pi$ MeV. The improvement with respect to the
 original paper in \cite{PICH} resides mainly in the hadronic
parametrization of the spectral function where, within the framework
of chiral Lagrangian and Weinberg sum rules, higher mass-state
contributions have been included into the spectral function.
Indeed, in the FESR approach, their role is important as
this sum rule is sensitive to the high-energy behaviour of the spectral
function. A priori, this is not the case of the Laplace sum rule (LSR),
as the exponential factor entering in it enhances
(when it is operative) the low-energy
contribution of the spectral function. For this reason, we might expect
that the Laplace sum rule will be less sensitive  to the higher mass
hadronic state effects, which are not under good control.
\section{ The four-quark two-point correlator}
The QCD expression of the spectral function has been evaluated in
\cite{PICH} including dimension-four operators, while the radiative
perturbative correction
has been re-evaluated recently \cite{JAMIN}. The RGI
two-point correlator reads:
\beq
\frac{1}{\pi} \mbox{Im}\hat{\Psi}
^{QCD}(t)= \frac{1}{(4\pi)^6}\frac{1}{10}
\frac{4}{3}t^4~\ga \alpha_s(t) \dr^{-4/9}
\aga 1-\frac{3649}{1620}\frac{\alpha_s}{\pi}
-40 \frac{\overline{m}^2_s}{t}- \frac{\Omega}{t^2}+...\adr,
\eeq
where:
\beq
\Omega= 200 \overline{m}^4_s-20\pi[16\pi m_s \la \bar ss \ra-
\la \alpha_s G^2 \ra ].
\eeq
We shall use the following values of the QCD parameters:
\beq
\Lambda_3=(0.35 \pm 0.10)~ \mbox{GeV},
\eeq
corresponding to $\alpha_s (M_Z) = 0.118 \pm 0.006$ to two-loop
accuracy. We shall use \cite{SNB}:
\beq
\hat{m}_s= (270 \pm 18)~\mbox{MeV}~~~~~~~~ \overline{m}_s= \hat{m}_s
\ga\frac{2\pi}{9\alpha_s }\dr^{-4/9} \ga 1+0.89
 \frac{\alpha_s}{\pi}+...\dr
\eeq
\beq
 \frac{\la \bar ss \ra}{ \la \bar dd \ra } \simeq 0.6-1~~~
{}~~~~~~~~\la \bar dd \ra = -\ga 188.3 ~\mbox{MeV}\dr^3
\ga\frac{2\pi}{9\alpha_s }\dr^{4/9},
\eeq
and:
\beq
\la \alpha_s G^2 \ra = (0.06 \pm 0.03)~\mbox{GeV}^4.
\eeq
The hadronic part of the spectral function will be parametrized in the
same way as in \cite{DOM}, which is given in Eq. (25) of that paper
(after correcting obvious misprints):
\bea
\frac{1}{\pi}\mbox{Im}\hat{\Psi}^{had}(t)&=&
|\hat{B}|^2\frac{t^2}{288\pi^2}\ga
\frac{f_K}{f_\pi}\dr^4
 . \int_{t_{10}}^{\ga\sqrt{t}-\sqrt{t_{20}}\dr^2}dt_1
\int_{t_{20}}^{\ga\sqrt{t}-\sqrt{t_1}\dr^2}dt_2 ~
\lambda^{1/2}\ga 1,\frac{t_1}{t},\frac{t_2}{t} \dr \nnb \\
&.&\Bigg\{ \ga \frac{t_1}{t}+\frac{t_2}{t}-1\dr^2
\frac{1}{\pi} \mbox{Im}\Pi^{(0)}(t_1)
\frac{1}{\pi} \mbox{Im}\Pi^{(0)}(t_2) \nnb  \\
&+&2\lambda\ga 1,\frac{t_1}{t},\frac{t_2}{t} \dr
\frac{1}{\pi} \mbox{Im}\Pi^{(1)}(t_1)
\frac{1}{\pi} \mbox{Im}\Pi^{(0)}(t_2) \nnb  \\
&+&\Bigg [\ga
\frac{t_1}{t}+\frac{t_2}{t}-1\dr^2 +8\frac{t_1t_2}{t^2}\Bigg]
\frac{1}{\pi} \mbox{Im}\Pi^{(1)}(t_1)
\frac{1}{\pi} \mbox{Im}\Pi^{(1)}(t_2) \Bigg\} \nnb \\
&+& \Theta(t-t_c)
\frac{1}{\pi}\mbox{Im}\Psi^{QCD}(t),
\eea
where the index $i$=0, 1
refers to the hadronic states with spin 0, 1, and:
\beq
\frac{1}{\pi} \mbox{Im}\Pi^{(i)}(t) \equiv
\frac{1}{\pi} \mbox{Im}\Pi^{(i)}_V(t) +
\frac{1}{\pi} \mbox{Im}\Pi^{(i)}_A(t),
\eeq
where the index $V,~A$ refers to the vector and axial-vector currents;
$\lambda^{1/2}$ is the usual phase space factor.
We shall use the parametrization with finite width corrections (FWC)
proposed in Ref. \cite{DOM}, but we shall compare the results with
the ones obtained in the narrow width approximation (NWA).
\section{The Laplace sum rule analysis of $B_K$}
We shall be concerned with the Laplace sum rule:
\beq
{\cal L} (\tau) = \int_{t_0}^{t_c} dt e^{-t\tau} \frac{1}{\pi}
\mbox{Im} \hat{\Psi}(t),
\eeq
where $t_0$ is the threshold specific for each channel, while $t_c$ is
the one of the QCD continuum coming from the QCD spectral function given
previously. The sum rule analysis is standard. We are looking for the
region of stability\footnote[1]{The meaning of
stability has been discussed in \cite{BELL} by comparing the exact
and approximate solutions of the sum rule for the harmonic oscillator
in quantum mechanics. What happens in QCD is very similar to this case.}
(minimum sensitivity)\footnote[2]{At the minima or at the
inflexion point, we have a balance between the continuum and the
nonperturbative contributions.},
in the sum rule variable $\tau$ and study the
$t_c$-dependence of this $\tau$-stability (stability in $t_c$ corresponds
to the minimum sensitivity in the higher meson mass contributions).
 For illustrating the analysis,
we show in Fig. 1 the $\tau$-stability for different choices of the QCD
parameters and for various values of $t_c$. It can be noticed that the
stability inflexion point is at $\tau=0.125$ GeV$^{-2}$, which is
exceptionally small compared with the hadronic scale 1 GeV$^{-2}$. This
feature indicates that the parametrization of the spectral function by
the $K\bar K$ alone will be a bad approximation, as expected from the
FESR analysis. This is essentially due to the quartic $t$-behaviour of
the
QCD spectral function and to the huge non-perturbative corrections, which
push to work at smaller $\tau$-values for the Operator Product Expansion
(OPE)
to make sense, but at this $\tau$-values, one
 is quite far away from the $K\bar K$ threshold.
However, at this small $\tau$-value, a possible large, but quantitatively
less controllable, as instanton-like effect appearing in high
powers of $\tau$ is $obviously$ negligible.
The result is shown in Fig. 2a (curve 0) where the value of
the $B$-parameter increases brutally with $t_c$. The situation is already
improved when one includes the $KK^*$ (curve 1), where a stability in
$t_c$ occurs at the inflexion point. We have used the parametrization of
the form factor including non-resonant $K\eta$ states proposed in
\cite{DOM}, which we have compared with the one within NWA using the $K^*
$-coupling $\gamma_{K^*}=(2.63\pm 0.16)$ from $\tau$-decay data and from
QCD spectral sum rule (QSSR)\cite{SNB}, with the normalization:
\beq
\frac{1}{\pi} \mbox{Im}\Pi^{K^*}(t) = \frac{M^2_{K^*}}{2\gamma^2_{K^*}}
\delta (t-M^2_{K^*}).
\eeq
The FWC is almost negligible although it
tends to pronounce the existence of the inflexion point.
Then, the positivity of the spectral
function implies an upper bound:
\beq
\Big[ f_K/(1.2f_\pi)\Big]^2\hat{B} \leq 0.83.
\eeq
We add to the former the $K^*K^*$ and $KK_1$ effects (curve 2). We use
the $K_1$-mass of 1.39 GeV, while we determine its coupling $\gamma_{K_1}
\simeq 4.6$ using the first Weinberg sum rule involving the $K^*,~K$,
$K_1$ and $K^*_0$ treated in a NWA.
We use $f_{K^*_0} \simeq (28.3 \pm 4.5)$ MeV (recall $f_\pi=93.3$ MeV)
from QCD spectral
sum rules (QSSR)\cite{SNB}. For the curve 3 (analogous to curve 3 of
\cite{DOM}), we add the effects of
$KK^*_0$ and $ K^*K_1$.
Although there is a big change from curve 1 to 2 (a similar
feature is obtained from FESR in Fig. 2 of \cite{DOM}), there is a
negligible effect from curve 2 to 3 (this is not the case of FESR). Here,
we see that the exponential factor starts to be operational as the
effects of the higher states become suppressed.
 Indeed, the inclusion of the
effects of the $K^*K^*_0$, $K^*K_1$, $K^*_0K^*_0$ and $K_1K_1$,
and some other combinations involving higher mass
states give negligible effect (at the optimization scale, the exponential
introduces a weight factor of about 0.3 instead of 1 for the FESR).
Therefore, we include them into the QCD continuum. The previous feature
is one big advantage of the Laplace sum rule compared with the FESR.
Also by comparing curve 2 here and in Fig. 2 of \cite{DOM}, we can
notice that the result from the LSR is more stable in $t_c$
than the one from FESR.

\nin
In Figs. 2,
we have an inflexion point at the value of $t_c=7.5-8$ GeV$^2$, which is
inside the duality region 6.5--9.5 GeV$^2$
expected from the FESR \cite{PICH,DOM}. We show
in Fig. 2b the effect of the subtraction point where the QCD parameters
are evaluated as well as the one due to the $SU(3)$-breaking of the
quark condensate. Taking into account different sources of uncertainties,
we deduce at the inflexion point:
\beq
\Big[ f_K/(1.2f_\pi)\Big]^2
\hat {B}_K = 0.55 \pm 0.02 \pm 0.05 \pm 0.05 \pm 0.05 ,
\eeq
where the errors are due respectively to the choice of the subtraction
point, the value of $\alpha_s$ or $\Lambda$, the SU(3)-breaking of the
condensate, and the localization of the inflexion point. Adding these
errors quadratically, we finally obtain the $best~estimate$:
\beq
\Big[ f_K/(1.2f_\pi)\Big]^2
\hat {B}_K = 0.55 \pm 0.09.
\eeq
A $conservative$ estimate could be obtained inside the whole range
of the $t_c$-$duality$ region from 6.5 to 9.5 GeV$^2$. In this case, one
obtains from Fig. 2b:
\beq
\Big[ f_K/(1.2f_\pi)\Big]^2
\hat {B}_K = 0.58 \pm 0.22,
\eeq
which, however, we expect to be a number with a too generous error.
On the other hand, by combining the results in Eqs. (18) and (19)
 with the one
in Eq. (5) (after rescaling the value of $f_K$) from the FESR, we deduce
the (weighted) $average$:
\beq
\Big[ f_K/(1.2f_\pi)\Big]^2
\hat {B}_K = 0.51 \pm 0.07.
\eeq
We have not included in this average the existing estimates from 3-point
function sum rules (see e.g. \cite{SNB} for a critical review on these
estimates). Indeed,
 due to the more complicated analytic property (in particular, we loose
the positivity property of the spectral function) of this
quantity, the inclusion of the higher resonance mass states into the
corresponding spectral function is therefore highly non-trivial
(interference terms) and can even become uncontrolled. In this respect,
we expect that the estimate from the 3-point function is less accurate
than from the 2-point function.
\section{Conclusions and values of the CKM parameters from the LSR}
We have derived the value of the invariant $B$-parameter using the
LSR. We have obtained the upper bound in Eq. (16), the
$best~estimate$ in Eq. (18) and the $conservative~estimate$ in Eq. (19).
Our result is slightly higher than the one from FESR, but agrees
with it within the errors. By combining the LSR and FESR estimates,
we deduce in Eq. (20) the $average$ from QSSR. Our previous result
agrees with the estimate  from the chiral
Lagrangian \cite{PRADES}. Our bound excludes  some other estimates
in the existing literature and raises again the question on the
$actual$ reliability of the value 0.8 from present
lattice calculations \cite{GUPTA}, which is, however, obtained
in the $SU(3)_f$ symmetry limit. In our approach, the $SU(3)_f$ symmetry
breaking is mainly reflected in the deviation of $f_K$ from $f_\pi$
(see Eq. (12)). Then, if we ignore the $SU(3)_f$ breaking,
i.e taking $f_K=f_\pi$, we would
obtain $\hat {B}_K \simeq 0.79$, in (surprising) good agreement with the
present lattice value.

\nin
Our estimate of $B_K$ combined with the
recent LSR estimates of $f_B$ \cite{SN1}, $B_B$\cite{PIVO}
and the value of the top mass will
certainly give stronger constraints on the CP-violation parameters
That can be explicitly seen
from various fits \cite{PRADES,ALI}.
Indeed, by using our $best$ estimate in Eq. (18) and our $best$
value \cite{SN1}:
\beq
f_B \sqrt{B_B}=(1.49\pm 0.14)f_\pi,
\eeq
which are,
actually, both obtained from only $one$ method (LSR for the two-point
functions), and
using the standard Wolfenstein parametrization,
one obtains for the CKM parameters of the unitarity triangle,
the best fit for $\chi^2=0.58$ \cite{ALI2}:
\beq
\rho \approx  0.09 ~~~~~~~~~~
\eta \approx 0.41,
\eeq
and, in Fig. 3, the present allowed domain of the set $(\rho,\eta)$.
We can consider the
previous set $(\rho,\eta)$ values as the $first$ consistent (in the
sense of a single type of sum rules used) estimate from QSSR.
One can also notice from \cite{ALI}
that $\rho$ is more sensitive to the changes
of the QCD input parameters than $\eta$.
\noindent
\section*{Figure captions}
{\bf Fig. 1:} $\tau$-stability of the invariant parameter $\hat{B}_K$
for different choices of the QCD parameters: {\bf a)} running mass and
coupling evaluated at $t$;
{\bf b)} the same as (a) but at $\tau$-variable;
{\bf c)} the same as (a) but at $\tau_0=0.125$ GeV$^{-2}$; {\bf d)} the
same as (b) but using $\la \bar ss \ra/\la \bar dd \ra=0.6$.
\vspace*{0.5cm}

\nin
{\bf Fig. 2a}: $t_c$-stability for different parametrizations of the
spectral function: {\bf curve 0}: $\bar KK$;
{\bf curve 1} $: (0)+KK^*$;
{\bf curve 2} : $(1)+KK_1+K^*K^*$; {\bf curve 3}: $(2)+K_1K^*
+KK^*_0$.
\vspace*{0.5cm}

\nin
{\bf {Fig. 2b:}} Effects of different QCD parameters
on the $t_c$-stability.
The label of each curve is the same as the one in Fig. 1.
\vspace*{0.5cm}

\nin
{\bf {Fig. 3:}} Allowed domain in the $\rho$-$\eta$-plane from
a simultaneous fit of the present data compiled in Table 1 of \cite{ALI}
and of the QCD parameters given in Eqs. (18) and (21). The triangle
represents the best fit.
\vfill \eject

\end{document}